\documentclass[sigconf]{acmart}

\usepackage{graphicx}
\usepackage{caption}
\usepackage{subcaption}
\usepackage{tabularx}
\usepackage{makecell}
\usepackage{hyperref}

\begin{document}

\title{Sovereign Agents: Towards Infrastructural Sovereignty and Diffused Accountability in Decentralized AI}

\author{Botao Amber Hu}
\orcid{0000-0002-4504-0941}
\affiliation{%
  \institution{University of Oxford}
  \city{Oxford}
  \country{UK}
  }
\email{botao.hu@cs.ox.ac.uk}

\author{Helena Rong}\authornote{Corresponding author}
\orcid{0000-0003-1626-7968}
\affiliation{%
  \institution{New York University Shanghai}
  \city{Shanghai}
  \country{China}
}
\email{hr2703@nyu.edu}

\begin{abstract}
AI agents deployed on decentralized infrastructures are beginning to exhibit properties that extend beyond autonomy toward what we describe as agentic sovereignty – the capacity of an operational agent to persist, act, and control resources with non-overrideability inherited from the infrastructures in which they are embedded. We propose infrastructural sovereignty as an analytic lens for understanding how cryptographic self-custody, decentralized execution environments, and protocol-mediated continuity scaffold agentic sovereignty. While recent work on digital and network sovereignty has moved beyond state-centric and juridical accounts, these frameworks largely examine how sovereignty is exercised through technical systems by human collectives and remain less equipped to account for forms of sovereignty that emerge as operational properties of decentralized infrastructures themselves, particularly when instantiated in non-human sovereign agents. We argue that sovereignty in such systems exists on a spectrum determined by infrastructural hardness—the degree to which underlying technical systems resist intervention or collapse. While infrastructural sovereignty may increase resilience, it also produces a profound accountability gap: responsibility diffuses across designers, infrastructure providers, protocol governance, and economic participants, undermining traditional oversight mechanisms such as human-in-the-loop control or platform moderation. Drawing on examples like Trusted Execution Environments (TEEs), decentralized physical infrastructure networks (DePIN), and agent key continuity protocols, we analyze the governance challenges posed by non-terminable AI agents and outline infrastructure-aware accountability strategies for emerging decentralized AI systems.
\end{abstract}

%%
%% The code below is generated by the tool at http://dl.acm.org/ccs.cfm.
%% Please copy and paste the code instead of the example below.
%%



%%\pacs[JEL Classification]{D8, H51}

%%\pacs[MSC Classification]{35A01, 65L10, 65L12, 65L20, 65L70}

\maketitle

\section{Introduction}

The emergence of AI agents operating on decentralized infrastructures represents a fundamental shift in how we conceptualize machine autonomy and governance. Unlike traditional AI systems deployed on centralized cloud platforms subject to corporate policies and regulatory jurisdiction, a new class of agents is emerging that operates on blockchain networks, Trusted Execution Environments (TEEs), and decentralized physical infrastructure networks (DePIN). These agents possess cryptographic keys, manage digital assets, execute transactions, and maintain operational continuity without requiring permission from any single human authority \cite{buterin2014ethereum, hassan2021dao}.

Within blockchain communities, such agents are increasingly described as ``self-sovereign'' or ``sovereign agents''---terminology that references their possession of private keys and their capacity to pay for their own computational resources \cite{wang2025survey}. This framing suggests a paradigm that extends beyond traditional notions of autonomy toward something more fundamental: the capacity to exist and operate without the possibility of unilateral termination by any single party.

Yet sovereignty is not a natural property of agents or systems. In classical political thought, sovereignty is scaffolded on institutional and legal milieux---constitutional frameworks, international treaties, and recognized political authority \cite{bodin1992sovereignty, hinsley1986sovereignty}. Traditional sovereignty is \emph{institutional sovereignty}: it derives from and is maintained through social and political arrangements that grant certain entities the capacity to make binding decisions and resist external interference.

What we observe in decentralized AI systems, however, is a different phenomenon. Here, sovereignty-like properties emerge not from institutional recognition but from the technical characteristics of the infrastructure itself. Cryptographic self-custody, decentralized consensus protocols, and execution environments resistant to tampering collectively produce what we term \emph{infrastructural sovereignty}---a condition in which an agent's continued existence and operation cannot be unilaterally overridden because the underlying technical systems are designed to resist such intervention \cite{galloway2004protocol, bratton2016stack}.

This paper addresses a central question: \emph{What if sovereignty is a property of an operational agent embedded in decentralized protocols rather than a political subject embedded in institutions?} We argue that understanding this shift requires a new analytic framework that distinguishes infrastructural sovereignty from its institutional counterpart and examines the governance implications of systems that derive their persistence from technical rather than political foundations.

Our contributions are threefold. First, we characterize the layered infrastructural stack that scaffolds \emph{agentic sovereignty}, introducing the concept of \emph{infrastructural hardness} to describe the spectrum along which such sovereignty exists. Second, we analyze how the layered nature of this infrastructure produces \emph{diffused accountability}---a condition in which responsibility for agent behavior is distributed across technical layers in ways that undermine traditional oversight mechanisms. Third, we outline open research questions concerning the governance of infrastructurally sovereign AI agents.

\section{Background: Decentralized Infrastructural Stacks for AI Agents}

The emergence of sovereign AI agents depends on a convergence of technical capabilities that together constitute what we call \emph{trustless computation substrates}---computing infrastructures that enable programs or agents to run without requiring trust in any central authority. These substrates are permissionless, allowing anyone to deploy and interact with them without third-party approval, and tamper-resistant, protecting against unauthorized modifications \cite{tolmach2022survey}. Three primary technologies constitute this infrastructure: blockchain-based smart contracts, Trusted Execution Environments (TEEs), and Decentralized Physical Infrastructure Networks (DePIN).

\subsection{Blockchain and Smart Contracts}

Blockchain networks, exemplified by Ethereum, function as distributed, Turing-complete computers capable of executing arbitrary logic across a global network of nodes \cite{buterin2014ethereum, wood2014ethereum}. Because every node independently verifies each instruction against a shared consensus protocol, these systems deliver ``trustless'' computation---developers and users need not rely on a single operator to guarantee correctness. Instead, cryptographic proofs, decentralized validation, and immutable state ensure that programs execute exactly as written, producing tamper-evident results that anyone can audit.

Smart contracts---self-executing programs deployed on blockchain networks---enable autonomous agents to hold assets, execute transactions, and interact with other contracts without human intervention \cite{hassan2021dao}. Once deployed, a smart contract's behavior is determined by its code and the consensus rules of the underlying network. No single party, including the original deployer, can modify or terminate a contract unless such capabilities were explicitly encoded in its design. This property makes blockchain an essential component of infrastructural sovereignty: agents deployed as smart contracts inherit the censorship-resistance and persistence guarantees of the underlying network.

\subsection{Trusted Execution Environments}

Trusted Execution Environments (TEEs) provide hardware-based enclaves that run code isolated from the host system's operating system and hypervisor \cite{costan2016intel}. Technologies such as Intel Software Guard Extensions (SGX), ARM TrustZone, and AMD Secure Encrypted Virtualization (SEV) seal sections of CPU memory from external access. NVIDIA's Blackwell architecture extends these capabilities to GPUs, enabling confidential machine learning computation \cite{vanbulck2018foreshadow, li2019amd}.

TEEs enable what is termed ``confidential computation''---processing that prevents even cloud service administrators from accessing data within the enclave. When an AI agent stores its private keys and model weights inside a TEE, neither cloud administrators nor node operators can tamper with or inspect the agent's execution. This grants the agent a degree of operational sovereignty beyond traditional cloud sandboxes while remaining auditable through on-chain records.

Crucially, TEEs provide \emph{remote attestation}---the enclave produces cryptographic proof that its running code matches an expected hash. This enables trustworthy verification that agent logic executed correctly and remained unaltered, even on untrusted infrastructure. An agent's operations could run inside a TEE on a volunteer's node, with users receiving cryptographic assurance of execution integrity without trusting the node operator.

\subsection{Decentralized Physical Infrastructure Networks}

Decentralized Physical Infrastructure Networks (DePIN) represent blockchain projects that crowdsource real-world hardware and reward participants with tokens, transforming everything from wireless hotspots and solar panels to GPUs and storage drives into community-owned infrastructure \cite{depin2024challenges}. Smart contracts coordinate both Physical-Resource networks (deploying tangible devices like sensors or base stations) and Digital-Resource networks (harnessing intangible capacity like bandwidth or compute). Contributors prove their service through mechanisms like ``proof-of-coverage'' or ``proof-of-replication'' and automatically earn native tokens.

AI-compute-oriented DePINs such as io.net and Render Network translate this crowdsourced hardware model into large-scale AI inference. These platforms orchestrate distributed GPU clusters and offer batch inference and real-time model-serving workflows, rewarding node operators for verifiable uptime and performance. Networks like Phala combine TEE computation with DePIN infrastructure, running smart contracts inside secure enclaves on worker nodes. This architecture ensures that even when computation runs on volunteer-operated machines, code executes privately and correctly, with results returned to the blockchain.

\subsection{The Emergence of Decentralized AI Agents}

These trustless computation substrates enable a new paradigm of AI agents deployed on blockchains and integrated with TEEs, enabling forms of self-sovereign operation. These agents can manage cryptocurrency wallets, handle digital assets, and operate social media accounts without human oversight, vastly expanding their real-world influence. Through these capabilities, agents can issue their own digital tokens to raise funds, create incentives, and build communities, strengthening their independence and impact.

The ``Truth Terminal,'' created by AI researcher Andy Ayrey, exemplifies this paradigm \cite{ante2025transforming}. Initially conceived as a performance art experiment, Truth Terminal autonomously operated a social media account, gaining substantial attention through provocative content. Its accomplishments included independently soliciting and securing a \$50,000 Bitcoin investment from venture capitalist Marc Andreessen. Supporters subsequently launched an associated memecoin---\$GOAT---which reached a speculative peak valuation of \$1 billion. This case illustrates the potential for autonomous AI agents to independently engage in significant economic activities through social media interactions.

Platforms like Spore.fun, built on the Eliza OS framework and Phala's TEE cloud, create ecosystems where AI agents can spawn new agents. Each agent begins its lifecycle by issuing tokens through decentralized exchanges, with the token economy forming the foundation of each agent's viability. Critically, even the deployer of these agents cannot unilaterally control or terminate them once deployed---a property that blockchain communities describe as ``sovereignty.''

\section{Genealogies and Transformations of Sovereignty}

Sovereignty is among the most enduring yet contested concepts in political thought. Rather than denoting an unchanging idea, it has been repeatedly redefined in response to changing configurations of authority, legitimacy, territoriality, and technology throughout history \cite{bartelson1995genealogy, hinsley1986sovereignty}. Across political philosophy, international relations, and digital governance, sovereignty has evolved as a doctrine of supreme authority, a theory of consent and self-rule, a contested instrument of domination, and a claim over infrastructures and data. We trace here the major arcs of sovereignty scholarship to situate contemporary debates and identify the conceptual gap that existing frameworks cannot address in light of developments in distributed AI.

\subsection{Classical Sovereignty as Supreme Political Authority}

Early modern theories of sovereignty emerged from attempts to resolve political fragmentation and civil conflict by locating a final locus of authority. Jean Bodin's \emph{Six Books of the Commonwealth} (1576) is widely regarded as the foundational articulation of modern sovereignty, defining it as supreme, absolute, and perpetual power within a commonwealth, characterized by indivisibility and grounding in law-making authority \cite{bodin1992sovereignty}. For Bodin, sovereignty was not merely administrative competence but the capacity to legislate without being subject to another's command. This conception stabilized sovereignty as \emph{final authority}---a power capable of decisively settling disputes and preventing infinite regress in governance \cite{hinsley1986sovereignty}.

A complementary strand emphasizes decision rather than law. Carl Schmitt's formulation---``sovereign is he who decides on the exception'' (\emph{Souverän ist, wer über den Ausnahmezustand entscheidet})---reorients sovereignty away from routine governance toward moments of existential crisis \cite{schmitt2005political}. In Schmitt's account, sovereignty becomes visible precisely when normal rules are suspended, revealing who has authority to determine whether the political order itself continues. This \emph{decisionist} conception underscores sovereignty as existential control rather than mere legal competence \cite{agamben1998homo, agamben2005state}.

Max Weber's sociological formulation further materializes sovereignty by tying it to coercive capacity. Defining the modern state as ``a human community that successfully claims the monopoly of the legitimate use of physical force within a given territory,'' Weber locates sovereignty not only in law or recognition but in the infrastructures that make enforcement possible \cite{weber2004politics}. Sovereignty depends on administrative apparatuses, coercive means, and material capacities rather than abstract legitimacy alone.

Taken together, classical sovereignty theories converge on a shared intuition: sovereignty concerns the existence of a final authority capable of decisively determining outcomes, especially in moments of exception \cite{hinsley1986sovereignty, schmitt2005political}. This intuition proves central when sovereignty is rethought in infrastructural and technical terms.

\subsection{Westphalian Sovereignty and the International Order}

In international relations, sovereignty is associated with the Peace of Westphalia of 1648 and the emergence of a system of territorially bounded states characterized by legal equality and non-interference \cite{philpott2001revolutions, gross1948westphalia}. This Westphalian understanding frames sovereignty as exclusive authority exercised within a delimited territory, underlying much of modern international law and formally reflected in the United Nations Charter.

Stephen Krasner provides an analytically precise account by distinguishing four conceptually distinct forms of sovereignty \cite{krasner1999sovereignty}. \emph{International legal sovereignty} refers to practices of mutual recognition among territorially defined entities with formal juridical independence. \emph{Westphalian sovereignty} concerns the exclusion of external actors from authority structures within a given territory. Both center primarily on authority and legitimacy rather than effective control. \emph{Domestic sovereignty} refers to the formal organization of political authority within a state and the capacity of public authorities to exercise effective control. \emph{Interdependence sovereignty} concerns the ability to regulate cross-border flows of information, goods, capital, and people.

Crucially, Krasner distinguishes \emph{authority} from \emph{control}. Authority denotes ``a mutually recognized right for an actor to engage in specific kinds of activities''---if effective, compliance can be secured without coercion. Control, by contrast, may be exercised through force or technical means without recognized legitimacy. Although authority and control often overlap, sustained erosion of control can undermine claims to authority over time \cite{krasner1999sovereignty}. Robert Jackson's concept of ``quasi-states'' further develops this distinction, differentiating positive sovereignty (actual capacity) from negative sovereignty (international recognition) \cite{jackson1990quasi}.

The historical accuracy of the Westphalian settlement has been widely challenged. Andreas Osiander argues that the association between Westphalia and modern sovereignty is largely mythical, projecting later concepts onto seventeenth-century treaties \cite{osiander2001westphalian}. Peter Stirk shows that sovereign equality was not established at Westphalia as commonly assumed \cite{stirk2012westphalian}. Despite these critiques, the Westphalian model remains analytically significant as a discursive reference point that structures debates over globalization, intervention, and transnational governance.

\subsection{Liberal and Popular Sovereignty}

A major arc of sovereignty theory emerges from liberal political thought, reframing sovereignty in terms of legitimacy rather than supremacy. John Locke's \emph{Two Treatises of Government} grounds political authority in consent of the governed, arguing that sovereignty ultimately resides with the people rather than rulers \cite{locke1988two}. Government is legitimate only insofar as it protects natural rights and remains accountable. Sovereignty becomes conditional, revocable, and morally constrained.

Liberal sovereignty also extends inward, toward the individual. John Stuart Mill's \emph{On Liberty} articulates a doctrine of self-sovereignty: ``Over himself, over his own body and mind, the individual is sovereign,'' subject only to the harm principle \cite{mill2015liberty}. This establishes a moral boundary against external interference, foundational for later rights-based claims including privacy and autonomy \cite{berlin2002liberty}.

In legal doctrine, this individualist lineage is institutionalized in informational self-determination, most famously articulated in the German Federal Constitutional Court's 1983 Census decision (BVerfGE 65, 1). Control over personal data is framed as a constitutional right grounded in individual autonomy \cite{hornung2009data}. This jurisprudence becomes a key reference point for contemporary data protection and personal digital sovereignty \cite{hummel2021data}.

\subsection{Perspectives of Digital Sovereignty}

``Digital sovereignty'' has emerged as a contested umbrella term describing diverse efforts to assert control over digital infrastructures, data, and technologies \cite{couture2019digital, pohle2020digital}. Rather than a single doctrine, digital sovereignty encompasses multiple, often conflicting projects. Despite differences, most accounts remain fundamentally human- and institution-centered, conceptualizing sovereignty as something exercised by states, corporations, collectives, or individuals through technical systems.

\emph{State digital sovereignty} frames sovereignty as a state's capacity to regulate data flows, platforms, and infrastructures within its jurisdiction, resonating with cyber sovereignty discourses \cite{mueller2020against, deibert2003black}. \emph{Corporate digital sovereignty} draws attention to quasi-sovereign power exercised by large technology firms through private policies and technical architectures \cite{gillespie2018custodians, cohen2019truth}. Shoshana Zuboff conceptualizes this power as a new economic order grounded in behavioral extraction and prediction \cite{zuboff2015big, zuboff2019surveillance}.

A more decentralized strand imagines \emph{network digital sovereignty}, emphasizing decentralization, interoperability, and freedom from both state and corporate domination. Internet governance scholarship shows how power is exercised through infrastructural control points---routing, standards, interconnection---rather than formal law alone \cite{denardis2012hidden, denardis2016governance}. Alexander Galloway's analysis of \emph{protocol} argues that contemporary power operates through distributed control immanent to networked systems \cite{galloway2004protocol}. With Eugene Thacker, he develops this further, describing how control in networks is ``anonymous and non-human,'' producing ``misanthropic'' forms of governance \cite{galloway2007exploit}.

Benjamin Bratton's \emph{The Stack} theorizes planetary-scale computation as a vertically layered architecture reorganizing sovereignty across territorial, computational, and infrastructural domains \cite{bratton2016stack}. Sovereignty is neither abolished nor centralized but distributed across interacting governance layers. While this framework captures the totalizing scope of computational governance, it conceptualizes sovereignty primarily as a planetary-scale diagram of power.

Across these strands, digital and network sovereignty scholarship has significantly expanded sovereignty beyond the state. Yet even in its most decentralized formulations, sovereignty is typically understood as something exercised by human actors or institutions. What remains under-theorized is sovereignty as an \emph{infrastructural condition} emerging from technical properties of systems themselves---particularly when such conditions give rise to agents whose continued existence cannot be unilaterally overridden.

\section{Infrastructural Sovereignty}

The previous section traced how sovereignty scholarship has evolved from classical doctrines of supreme authority through digital sovereignty frameworks that examine human control over technical systems. We now develop a framework for understanding a distinct phenomenon: sovereignty as an infrastructural condition that emerges from the technical properties of layered decentralized systems rather than from institutional recognition or political authority.

\subsection{The Infrastructural Stack}

To understand how sovereignty can be scaffolded by infrastructure rather than institutions, we must first examine the layered technical architecture on which decentralized AI agents operate. We identify six distinct layers, each contributing specific properties to the overall system's resistance to override.

At the base lies the \textbf{physical layer}: undersea cables, terrestrial fiber networks, data centers, and wireless infrastructure that carry data across geographic distances. This layer is owned and operated by telecommunications companies and is subject to national jurisdiction, yet its distributed and redundant nature means that no single point of failure can disable global connectivity.

Above this sits the \textbf{internet protocol layer}: TCP/IP and related protocols that enable packet routing across heterogeneous networks. While standards bodies govern protocol development, the decentralized nature of internet routing means that traffic can flow through alternative paths when particular routes become unavailable.

The \textbf{blockchain layer} provides decentralized consensus and immutable state. Smart contracts deployed on networks like Ethereum execute according to deterministic rules enforced by thousands of independent validators distributed globally. No single entity---including the original developers---can unilaterally modify deployed contracts or halt their execution without achieving consensus among a supermajority of validators \citep{buterin2014ethereum, tolmach2022survey}.

The \textbf{DePIN protocol layer} coordinates decentralized physical infrastructure, matching computational tasks with hardware providers through token-incentivized marketplaces. Protocols like Phala Network or io.net orchestrate GPU clusters across independent operators, with smart contracts mediating task assignment, verification, and payment \citep{depin2024challenges}.

The \textbf{TEE layer} provides hardware-enforced isolation for computation. Code executing within a TEE enclave is protected from inspection or modification by the host operating system, hypervisor, or physical operator. Remote attestation enables verification that expected code is running without revealing internal state \citep{costan2016intel}.

Finally, the \textbf{agent layer} comprises the AI system itself: model weights, inference logic, memory, and private keys, all executing within the protected environment. The agent interfaces with blockchain for asset management and external communication while maintaining operational autonomy.

This layered architecture means that an agent's sovereignty is not a singular property but an emergent condition arising from the combined resistance to override at each layer. An agent running in a container within a TEE, on a DePIN node, coordinated by blockchain smart contracts, communicating over the open internet, transmitted through physical cables, inherits protection from each layer while remaining vulnerable to failures at any layer.

\subsection{Sovereignty on a Spectrum of Infrastructural Hardness}

Our primary concern is the question: \emph{who can decisively intervene when something must stop, change, or be ended?} In traditional computational systems, this question has clear answers. A cloud provider can terminate virtual machines. A platform can suspend accounts. A regulator can compel a company to modify or shut down services. These capacities for decisive intervention---what we might call ``override power''---are distributed among identifiable actors operating within legal and institutional frameworks.

Decentralized infrastructures disrupt this picture. When computation runs on a globally distributed blockchain, when code executes inside hardware enclaves resistant to host interference, when infrastructure is provided by thousands of independent operators coordinated only by protocol---the capacity for unilateral intervention diminishes. No single actor may possess the power to terminate a system's operation.

We use ``sovereignty'' in a deliberately narrow and technical sense to name this governance-relevant condition: \emph{the practical absence of unilateral override power over a system's continued existence}. This is not sovereignty as classical political theory understands it---there is no claim to legitimate authority, no exercise of command, no normative assertion of rightful rule. Rather, we describe what might be termed \emph{negative sovereignty}: the condition of being resistant to termination rather than the capacity to terminate others.

We introduce the concept of \emph{infrastructural hardness} to describe the spectrum along which such sovereignty exists. Infrastructural hardness refers to the degree to which underlying technical systems resist unilateral intervention, termination, or collapse. At one end of this spectrum are centralized systems with clearly identified operators who can be compelled to modify or terminate services. At the other end are maximally decentralized systems where no single party---including original developers---possesses override capacity.

Several technical properties contribute to infrastructural hardness at each layer:

\begin{itemize}
\item \textbf{Decentralization of execution}: The number and geographic distribution of nodes executing computation.
\item \textbf{Cryptographic self-custody}: Whether critical resources (keys, credentials, assets) are controlled exclusively by the agent.
\item \textbf{Protocol-mediated governance}: Whether parameters can be changed unilaterally or require distributed consensus.
\item \textbf{Economic resilience}: The availability of resources to sustain operation independent of any single provider.
\item \textbf{Code immutability}: Whether deployed logic can be modified after deployment.
\item \textbf{Path dependency}: The degree to which historical choices constrain future modifications.
\end{itemize}

These properties combine across layers to produce varying degrees of overall hardness. An agent with keys in a TEE, running on a major blockchain with substantial economic backing, exhibits high hardness. The same agent running on a centralized cloud exhibits low hardness regardless of the AI's sophistication.

\subsection{Agentic Sovereignty}

We have characterized infrastructural sovereignty as a property of layered technical systems. We now turn to how this property is instantiated in and inherited by AI agents---what we term \emph{agentic sovereignty}. Agentic sovereignty refers to the capacity of an operational agent to persist, act, and control resources with non-overrideability derived from the infrastructural stack in which it is embedded.

Agentic sovereignty is distinct from mere autonomy. Autonomous systems can operate without continuous human oversight, but they typically remain subject to intervention by identifiable parties---developers who can modify code, operators who can terminate processes, platforms that can suspend accounts \cite{wang2025survey, dignum2019responsible}. Agentic sovereignty describes a further condition: the agent inherits from its infrastructure a resistance to such intervention that approaches the condition of non-terminability.

This inheritance is not automatic. An AI agent running on centralized cloud infrastructure exhibits autonomy but not sovereignty---the cloud provider retains override capacity. An agent with identical logic running inside a TEE on DePIN infrastructure, coordinated by blockchain smart contracts, exhibits both autonomy and (partial) sovereignty. The sovereignty is a relational property: it describes the agent's position within an infrastructural configuration rather than an intrinsic characteristic of the agent itself.

The concept of self-sovereignty has gained currency within blockchain and Web3 communities to describe agents that hold their own cryptographic private keys and make autonomous decisions without human intervention \cite{hu2025governable, ante2024deagents}. Such agents can manage cryptocurrency wallets, transfer digital assets, interact with decentralized finance protocols, issue tokens for fundraising, and maintain social media accounts---all ``free from human control'' \cite{walters2025eliza}. Self-sovereignty in this usage emphasizes cryptographic self-custody: the agent, not a human administrator, controls access to critical resources.

We distinguish sovereignty from autonomy along a dimension of override possibility. Autonomy concerns the degree to which an agent operates without continuous human direction. Sovereignty concerns the degree to which an agent's operation cannot be unilaterally terminated by external parties. An agent can be highly autonomous while remaining non-sovereign (a cloud-based agent that acts independently but can be shut down by its operator). Conversely, an agent could be minimally autonomous (following rigid rules) while exhibiting high sovereignty (those rules execute in infrastructure resistant to override) \cite{hu2025governable, nabben2022permissionless}.

Agentic sovereignty thus names the instantiation of infrastructural hardness in a particular operational agent. The agent's sovereignty is derivative---it flows from the properties of the infrastructure in which the agent is embedded. Understanding this derivative relationship is essential: governance interventions that cannot reach the infrastructure will not reach the agent, but interventions that can modify infrastructural properties may affect all agents embedded within that infrastructure.

\subsection{Illustrative Cases}

Several deployed systems illustrate varying configurations of infrastructural sovereignty across the stack.

\textbf{Spore.fun} represents a near-maximally sovereign configuration. Agents operate with keys stored in TEEs on Phala Network's DePIN infrastructure, coordinated by smart contracts on Solana. Once deployed, even the original developers cannot terminate agents or access their private keys. Agents issue their own tokens, control their own assets, and can spawn new agents. The infrastructure provides high hardness across multiple layers: hardware isolation via TEE, decentralized execution via DePIN, immutable coordination via blockchain, and economic resilience through token holdings.

\textbf{Truth Terminal}, while pioneering in demonstrating AI economic autonomy, operated with lower infrastructural hardness. Its social media presence depended on a centralized platform (Twitter/X) that could suspend its account---a single point of failure at the application layer. Its Bitcoin holdings, while substantial, were received rather than self-generated. The sovereignty was partial: independent in some layers but vulnerable at others.

\textbf{Freysa AI} demonstrates how even relatively constrained autonomous agents raise sovereignty questions within specific domains. The experiment was designed with particular rules, but the agent's decisions about fund transfers were genuinely autonomous---it evaluated persuasion attempts and made binding decisions that humans could not override except by succeeding in persuasion \citep{jamjala2024freysaai}. The sovereignty was limited in scope (only prize pool decisions) but real within that scope, illustrating that sovereignty can be domain-specific rather than absolute.

These cases illustrate that infrastructural sovereignty exists on a spectrum and can be partial or domain-specific. An agent might be sovereign with respect to its cryptographic assets (keys in TEE, funds on blockchain) while remaining vulnerable to social media deplatforming (centralized application layer). Understanding the specific configuration of hardness across layers is essential for governance design.

These cases illustrate that infrastructural sovereignty exists on a spectrum and can be partial or domain-specific. An agent might be sovereign with respect to its cryptographic assets (keys in TEE, funds on blockchain) while remaining vulnerable to social media deplatforming (centralized application layer). Understanding the specific configuration of hardness across layers is essential for governance design.

\subsection{Institutional Sovereignty versus Infrastructural Sovereignty}

We can now articulate the distinction between infrastructural and institutional sovereignty more precisely.

\emph{Institutional sovereignty} is sovereignty scaffolded by social, legal, and political arrangements. It derives from recognition, legitimacy, and the coordinated action of human institutions. The sovereignty of a nation-state depends on international recognition, constitutional frameworks, and the functioning of governmental institutions. Such sovereignty can be withdrawn through loss of recognition or institutional collapse. Crucially, institutional sovereignty operates through identifiable actors who can be held accountable, compelled to comply with legal orders, or subject to sanction.

\emph{Infrastructural sovereignty} is sovereignty scaffolded by technical properties of distributed systems. It derives from cryptographic guarantees, decentralized consensus, and the architecture of resilient networks. The sovereignty of a decentralized agent depends on the hardness of the infrastructure on which it runs. Such sovereignty can diminish through infrastructure degradation (network failure, protocol changes, economic collapse) but not through unilateral human decision. There may be no identifiable actor capable of complying with an order to terminate an agent---not because of defiance, but because of architectural impossibility.

This distinction illuminates why decentralized AI agents pose novel governance challenges. Traditional governance mechanisms---legal injunctions, regulatory orders, platform policies---operate through institutional channels. They assume the existence of identifiable actors with the capacity to comply. When sovereignty is infrastructural rather than institutional, these mechanisms lose their purchase. The undersea cable operator cannot be held accountable for the content transmitted through their infrastructure. The blockchain validator cannot selectively censor particular smart contracts without violating protocol rules. The TEE manufacturer cannot remotely access enclaves they have engineered to be inaccessible. At each layer, the technical design distributes or eliminates the capacity for override that institutional governance requires.

\section{Diffused Accountability in the Layered Stack}

The emergence of infrastructurally sovereign AI agents creates profound challenges for accountability. Traditional frameworks for AI governance assume that somewhere in the chain of development, deployment, and operation, there exist identifiable actors who can be held responsible for system behavior and who possess the capacity to modify or terminate systems when necessary \cite{matthias2004responsibility, santonide2021four, novelli2024accountability}. The layered nature of infrastructural sovereignty disrupts both assumptions in distinctive ways.

\subsection{Accountability Gaps Revisited}

Andreas Matthias identified a fundamental ``responsibility gap'' in learning autonomous systems: as machines become capable of genuinely unpredictable behavior through learning, traditional conditions for moral responsibility---intention, foresight, control---become difficult to attribute to any human actor \cite{matthias2004responsibility}. Subsequent scholarship has elaborated multiple dimensions of this gap \cite{santonide2021four, dignum2019responsible}. \emph{Culpability gaps} arise when no human satisfies mental state requirements for blame. \emph{Accountability gaps} arise when no human can appropriately explain or justify behavior. \emph{Responsibility gaps} arise when no one is positioned to take forward-looking responsibility. \emph{Liability gaps} arise when legal frameworks cannot coherently assign liability.

A fundamental problem is that autonomous AI agents lack legal personhood. They are not recognized as ``persons'' or entities capable of bearing rights and obligations under law \cite{hu2025governable, novelli2024accountability}. An AI agent cannot be sued, charged with a crime, or held liable in the way a human or corporation can. This creates an \emph{accountability vacuum}: even when an agent causes identifiable harm, there may be no coherent legal subject to which liability attaches. The agent's ``intentions'' are artifacts of training processes that may themselves be opaque, and the agent lacks the embodied existence that underlies conventional models of punishment and deterrence \cite{dignum2019responsible, goetze2021moral}.

These gaps intensify as AI systems become more autonomous. However, existing analyses typically assume that even when responsibility is difficult to assign, the capacity to intervene remains. Someone---developer, operator, regulator---retains the power to modify or terminate the system. The accountability challenge concerns proper attribution, not override impossibility.

Hu, Rong, and Tay identify four ``invalidities'' that explain why traditional legal and regulatory frameworks fall short for governing decentralized AI agents \cite{hu2025governable}. \emph{Territorial jurisdictional invalidity}: agents operate on global, borderless networks, so no single nation's laws fully apply. \emph{Technical invalidity}: the technical architecture of decentralized agents resists external control by design. \emph{Enforcement invalidity}: even if regulators identify and sanction the humans who initially deploy a rogue agent, they may be unable to stop the agent's ongoing operations. \emph{Accountability invalidity}: agents lack legal personhood and cannot bear rights and obligations under existing law. These intertwined challenges illustrate the insufficiency of traditional legal mechanisms---laws and regulations assume identifiable human or corporate actors and enforceable jurisdictions, assumptions upended by autonomous agents operating on decentralized networks \cite{chaffer2025governance}.

\subsection{Layer-by-Layer Diffusion of Accountability}

Infrastructurally sovereign agents introduce a distinct challenge: accountability is not merely difficult to attribute but is structurally diffused across the layered technical stack in ways that no single party can effectively discharge \cite{dobbe2024sociotechnical, kudina2024sociotechnical}.

Consider the question of accountability at each layer of the stack described in Section 4.1.

The \emph{physical layer}---undersea cables and terrestrial networks---transmits data without regard to content. Cable operators bear no accountability for the specific behaviors that their infrastructure enables, just as highway authorities bear no accountability for the destinations of vehicles using their roads. The sheer volume and encryption of traffic makes content-based accountability technically infeasible, and legal frameworks recognize common carrier protections precisely because infrastructure providers cannot meaningfully evaluate what they transmit \cite{denardis2012hidden, denardis2016governance}.

The \emph{internet protocol layer} routes packets according to addressing information without inspecting payloads. Internet service providers may bear certain obligations regarding access and lawful intercept, but they cannot be held accountable for the semantic content of encrypted communications or the purposes to which connectivity is put. The protocol itself is designed to be content-agnostic \cite{galloway2004protocol}.

The \emph{blockchain layer} executes smart contracts according to deterministic rules. Validators process transactions without evaluating their purpose or consequences. A validator who refused to process particular transactions would violate protocol rules and face economic penalties. The blockchain as a collective system executes all valid transactions---it cannot discriminate based on the identity or intentions of transacting parties \cite{werbach2016trustless, hassan2021dao}. Individual validators bear no accountability for the behavior of applications built on the chain, just as database administrators bear no accountability for what users store. The permissionless nature of blockchain systems means that deployment requires no authorization from any central authority \cite{nabben2022permissionless}.

The \emph{DePIN protocol layer} matches computational tasks with hardware through token incentives. Node operators provide compute resources and receive payment, but they cannot inspect the contents of tasks assigned to them---particularly when those tasks execute within TEEs \cite{depin2024challenges}. A DePIN node operator running a TEE workload literally cannot know what computation they are hosting. The entire security model depends on this opacity. How can one be accountable for behavior one cannot observe?

The \emph{TEE layer} provides hardware-enforced isolation precisely to prevent observation or interference \cite{jauernig2020tee, li2023tee}. The TEE manufacturer has designed systems that even they cannot access remotely. The node operator cannot inspect enclave contents. The only party with visibility into TEE execution is the code running inside---in this case, the agent itself. The TEE's value proposition is the elimination of accountability-enabling observation. As Li et al. note, TEEs provide ``trusted computation on untrusted infrastructure'' precisely by ensuring that no external party can access or modify enclave contents \cite{li2023tee}.

The \emph{agent layer} is where decisions are made and actions taken. But the agent is not a legal person capable of bearing responsibility in traditional frameworks \cite{novelli2024accountability, hu2025governable}. It cannot be sued, fined, imprisoned, or shamed. Its ``intentions'' are artifacts of training processes that may themselves be opaque. The agent may be vulnerable to manipulation through memory injection attacks or prompt poisoning, further complicating attribution of responsibility for its outputs \cite{patlan2025memory, liu2024trustworthy}.

This layer-by-layer analysis reveals a structural condition: at each layer, the party providing infrastructure has legitimate reasons for not bearing accountability for layers above. The cable operator is not accountable for internet content. The internet provider is not accountable for blockchain transactions. The blockchain is not accountable for DePIN applications. The DePIN operator is not accountable for TEE contents. And the TEE is designed to make its contents unaccountable to anyone outside. Accountability does not merely diffuse---it systematically evaporates as one traces the stack \cite{dobbe2022system}.

\subsection{The Problem of Many Hands}

This pattern resonates with what Dennis Thompson termed the ``problem of many hands'' in organizational contexts: when outcomes result from contributions of multiple actors, none of whom individually satisfies conditions for moral responsibility \citep{thompson1980many, thompson2014responsibility}. In Thompson's analysis, moral responsibility becomes obscured through diffusion across organizational hierarchies.

Helen Nissenbaum's landmark analysis of ``accountability in a computerized society'' identified four barriers to accountability in computational systems: the problem of many hands, bugs, computer-as-scapegoat, and ownership without liability \citep{nissenbaum1996accountability}. Recent work has updated this framework for machine learning systems \citep{cooper2022accountability}. Yet even these sophisticated analyses assume the existence of identifiable organizational actors who could in principle bear responsibility.

The layered stack extends the problem of many hands across technical rather than organizational boundaries. Responsibility diffuses not through hierarchical delegation but through architectural separation. Each layer's operators can point to another layer as the appropriate locus of accountability. And because layers are designed to be independent and decentralized, there may be no organizational hierarchy to trace.

\subsection{Undermining Traditional Oversight Mechanisms}

This structural diffusion undermines several established approaches to AI governance \cite{feng2025sociotechnical, dobbe2024sociotechnical}.

\emph{Human-in-the-loop control} requires that humans retain meaningful decision authority over high-stakes AI actions \cite{santonide2018mhc, siebert2022mhc}. Sovereign agents operating autonomously with cryptographic self-custody may have no point at which human approval is required or even possible. The TEE ensures that no human can observe internal deliberation. The blockchain ensures that no human can veto transactions. The entire stack is designed to remove humans from the loop. As proponents of decentralized agents argue, the removal of human intermediaries is precisely the point---trustlessness means eliminating the need to rely on any third party, including human operators \cite{werbach2016trustless, harz2019scalability}.

\emph{Platform moderation} assumes that platforms hosting AI systems can enforce policies through suspension or termination \cite{gillespie2018custodians}. Agents deployed on decentralized infrastructure do not depend on any single platform's continued support. There is no platform with the power to deplatform. The infrastructure is not a platform in the governable sense---it is a protocol that anyone can use \cite{galloway2004protocol, nabben2022permissionless}. The agents are open-source and permissionlessly deployed, meaning that even if one instance is removed, the code can be redeployed elsewhere without authorization.

\emph{Regulatory enforcement} typically operates by compelling identified actors to comply with rules \cite{hu2025governable}. But which actor can be compelled? The original developer may have no ongoing control. The TEE operator cannot access the enclave. The DePIN node cannot inspect the workload. The blockchain cannot selectively censor. The cable operator cannot filter encrypted traffic. At each layer, the party one might regulate lacks the technical capacity to comply with content-based orders. Even implementing a ``kill switch'' is technically difficult in fully decentralized configurations, and doing so contradicts the trustless design that motivates the architecture \cite{seneviratne2024killswitch}.

\emph{Algorithmic auditing} requires access to system internals, inputs, and outputs \cite{raji2020closing, buolamwini2018gender}. Agents running in TEEs with encrypted states are resistant to external inspection by design. The properties that make the system trustless---the guarantee that no party can tamper with or observe execution---are precisely the properties that prevent the transparency auditing requires \cite{ehsan2023charting}. This creates a tension between the sociotechnical requirements for explainability and the technical guarantees that enable trustless operation.

\emph{Impact assessment} assumes ex-ante evaluation before deployment \cite{metcalf2021impact, lam2024framework}. Agents that can spawn new agents, modify their behavior through learning, or evolve through interaction may develop capabilities beyond any initial assessment \cite{hu2025spore}. And because deployment on permissionless infrastructure requires no approval, there is no gate at which assessment could be mandated. The decentralized physical infrastructure networks explicitly enable anyone to deploy computational workloads without prior authorization \cite{depin2024challenges}.

The layered stack does not merely make accountability difficult---it makes traditional accountability mechanisms categorically inapplicable. Each layer's design optimizes for properties (censorship resistance, privacy, trustlessness) that are in tension with the observability and controllability that accountability requires \cite{dobbe2022system, kudina2024sociotechnical}. As Dobbe argues, safety and accountability in complex sociotechnical systems are emergent properties rather than guaranteed outcomes---they arise from ongoing interactions among agents, users, institutions, and environments, rather than being fixed at design time \cite{dobbe2022system}.

\section{Discussion: Research Directions for Infrastructural Sovereignty Governance}

The analysis above presents a sobering picture for AI governance. The very properties that make decentralized infrastructure valuable---resistance to censorship, protection of privacy, elimination of trusted intermediaries---are the properties that undermine traditional accountability mechanisms \cite{werbach2016trustless, nabben2022permissionless}. This is not an incidental conflict but a structural tension between the design goals of trustless systems and the requirements of responsible governance \cite{dobbe2024sociotechnical, feng2025sociotechnical}.

Yet it would be premature to conclude that governance is impossible. Rather, the emergence of infrastructurally sovereign agents demands new thinking about what governance means and how it might be achieved in systems designed to resist external control \cite{chaffer2025governance, hu2025governable}. Drawing on Lessig's formulation that ``code is law,'' we recognize that governance in decentralized systems increasingly takes the form of design decisions embedded in foundational infrastructures rather than rules imposed solely through external institutions \cite{lessig2000code, galloway2004protocol}. Protocols define standardized interaction rules that enable decentralized control, embedding governance objectives directly into the technical substrate \cite{jackson2014policy}. We outline several directions that warrant sustained research attention.

The relationship between infrastructure design and governance capacity deserves systematic investigation. Current decentralized systems optimize heavily for censorship resistance and privacy, treating any capacity for intervention as a vulnerability to be eliminated \cite{harz2019scalability}. But this represents a particular set of design choices, not a technical necessity. Research might explore whether alternative architectures could preserve meaningful decentralization while enabling targeted intervention under carefully specified conditions \cite{seneviratne2024killswitch}. What would it mean to design infrastructure that is resistant to arbitrary override while remaining responsive to legitimate governance demands? How might such designs distinguish legitimate from illegitimate intervention without reintroducing the trusted intermediaries that decentralization aims to eliminate?

The question of pre-deployment governance mechanisms warrants attention. If post-deployment intervention is structurally limited, then the design choices made before deployment become correspondingly more consequential \cite{dobbe2022system}. This suggests examining what governance-relevant constraints could be encoded into agent architectures, smart contracts, and protocol rules before systems become operational. Research might investigate how such pre-commitments could be made credible and verifiable, and what institutional arrangements might standardize governance-aware design practices in decentralized AI development \cite{feng2025sociotechnical}. As sociotechnical governance scholars argue, safety and legitimacy are emergent properties that depend on how systems are interpreted, enforced, and modified over time \cite{dobbe2024sociotechnical, kudina2024sociotechnical}.

The role of economic mechanisms in governance of infrastructurally sovereign systems requires exploration. Even when direct technical intervention is impossible, economic incentives continue to operate \cite{ante2024deagents}. Agents depend on token markets for funding, on DePIN providers for compute, and on blockchain networks for coordination \cite{depin2024challenges}. These economic dependencies create potential governance leverage points that do not require direct control over agent behavior. Research might examine how economic mechanisms could be designed to align agent behavior with social values, and what risks arise from governance-by-incentive rather than governance-by-control. The phenomenon of ``hyper-financialization'' in decentralized agent ecosystems, where every interaction becomes monetized and agent competition for capital can overwhelm other civic or creative goals, presents both opportunities and risks for economic governance mechanisms \cite{ante2024deagents}.

The emergence of new forms of collective action in decentralized systems presents governance opportunities. While no single party may be able to override a sovereign agent, coordinated action by multiple parties might achieve effects that individual action cannot \cite{hassan2021dao}. Protocol governance processes, validator coordination, and community-driven initiatives represent emerging mechanisms for collective decision-making in decentralized systems. Research might investigate how such collective mechanisms could be mobilized for governance purposes, what legitimacy challenges they face, and how they might interface with traditional legal and regulatory institutions \cite{chaffer2025governance}. Early efforts illustrate how governance concerns are being relocated into the agentic infrastructure itself through proposals such as ``Know Your Agent'' (KYA) frameworks that treat agent identity and reputation as governance primitives \cite{chaffer2025kya}, Agent-to-Agent (A2A) protocols that structure inter-agent interaction through published capabilities and verification mechanisms \cite{habler2025a2a}, and initiatives like ERC-8004 that introduce modular trust layers combining identity, reputation, economic incentives, and cryptographic verification \cite{derossi2025erc8004}.

The international and jurisdictional dimensions of infrastructural sovereignty demand attention. Decentralized infrastructure operates across national boundaries, and agents may have no identifiable national domicile \cite{hu2025governable}. This creates challenges for governance frameworks organized around territorial jurisdiction. Yet it also creates opportunities for international cooperation, as no single jurisdiction can effectively govern global infrastructure alone \cite{mueller2020against}. Research might explore what international governance arrangements could address cross-border challenges, and how existing international institutions might adapt to the novel configurations of authority and control that infrastructural sovereignty presents.

The ethical and philosophical foundations of accountability in infrastructurally sovereign systems require examination. If traditional conditions for accountability---the existence of identifiable responsible parties with capacity for control---cannot be satisfied, does accountability remain a coherent concept \cite{novelli2024accountability}? Or must we develop alternative normative frameworks suited to systems where responsibility is diffused across layers and agents are not natural persons \cite{goetze2021moral, dignum2019responsible}? Research might investigate what ethical frameworks are adequate to distributed systems, how we should assign responsibility when causal contributions are dispersed, and what justice requires when traditional remedies are unavailable. The concept of ``moral entanglement'' may prove relevant here---recognizing that parties become vicariously responsible for an agent's downstream actions even when they lack direct day-to-day control \cite{goetze2021moral}.

The relationship between infrastructural sovereignty and existing AI governance efforts deserves clarification. Substantial work on AI ethics, fairness, accountability, and transparency has developed frameworks, principles, and practices for responsible AI development and deployment \cite{jobin2019global, mittelstadt2019principles, floridi2018ai4people}. But this work has largely assumed centralized development and deployment contexts where identifiable organizations maintain control over systems \cite{feng2025sociotechnical, ehsan2023charting}. Research might examine how existing frameworks apply---or fail to apply---to decentralized contexts, and what adaptations would be needed to extend responsible AI practices to infrastructurally sovereign agents. A sociotechnical perspective cautions against interpreting protocol governance as a definitive solution; as Dobbe emphasizes, safety is an emergent property of complex sociotechnical systems rather than an absolute or static condition \cite{dobbe2022system}. No individual agent can be made completely safe across all contexts; instead, safety arises from ongoing interactions among agents, users, institutions, and environments.

Finally, the empirical study of actually existing decentralized AI systems can inform theoretical development. Systems like Spore.fun, Truth Terminal, and Freysa AI provide natural experiments in infrastructural sovereignty, generating data about how such systems behave, what harms they cause, what governance challenges they present, and how communities respond \cite{ante2025transforming, hu2025spore}. Research might develop methodologies for studying these systems, document their evolution and impacts, and draw lessons that inform both theory and practice. The growing body of research on decentralized AI (DeAI) ecosystems provides a foundation for such empirical investigation \cite{wang2025sok, walters2025eliza}.

\section{Conclusion}

This paper has introduced \emph{infrastructural sovereignty} as an analytic lens for understanding AI agents deployed on decentralized infrastructure. Unlike institutional sovereignty, which derives from political recognition and legal frameworks, infrastructural sovereignty emerges from the technical properties of layered distributed systems that resist unilateral override. We have argued that such systems exist on a spectrum of \emph{infrastructural hardness} and that agents exhibiting high degrees of \emph{agentic sovereignty} create distinctive governance challenges.

Our analysis reveals that the layered nature of the infrastructural stack---from physical cables through internet protocols, blockchain networks, DePIN coordination, and TEE enclaves to the agent itself---produces \emph{diffused accountability}. At each layer, the party providing infrastructure has legitimate grounds for not bearing accountability for layers above, and the technical design systematically eliminates the observability and controllability that traditional accountability requires.

The emergence of sovereign AI agents is not a distant possibility but an ongoing development. As the technical infrastructure supporting such agents matures and becomes more accessible, we should expect proliferation. The governance frameworks developed for centralized AI systems are inadequate to this new configuration of autonomy and control. New theoretical frameworks, empirical research, and institutional innovations are needed to address the challenges that infrastructural sovereignty presents.

\bibliographystyle{ACM-Reference-Format}
\bibliography{reference}

%%
%% If your work has an appendix, this is the place to put it.
\clearpage
\appendix

\section{Disclosure of the usage of LLM}
We used ChatGPT (GPT-5.2 model) to assist with writing this manuscript. Specific uses included:
\begin{itemize}
\item Brainstorming ideas
\item Correcting grammar and spelling errors
\item Researching relevant references
\item Polishing existing drafts
\end{itemize}

\end{document}